\documentclass[10pt,twocolumn]{article} %% two column, final layout

\usepackage{ol}
\usepackage{hyperref}
\usepackage{amsmath}

\begin{document}

\twocolumn[ %% activate for two-column option

\title{Strong optical force induced by morphology dependent resonances}

\author{Jack Ng, C.T. Chan, and Ping Sheng}
\address{Department of Physics, Hong Kong University of Science and Technology, Clearwater Bay, Hong Kong, China}
\author{Zhifang Lin}
\address{Department of Physics, Fudan University, China}

% Do not use \email or \homepage here. E-mail and URL can be given just before references.

\begin{abstract}
We consider the resonant optical force acting on a pair of transparent 
microspheres by the excitation of the Morphology Dependent Resonance (MDR). 
The bonding and anti-bonding modes of the MDR correspond to strong 
attractions and repulsions respectively. The dependence of the force on 
separation and the role of absorption are discussed. At resonance, the force 
can be enhanced by orders of magnitude so that it will dominate over other 
relevant forces. We find that a stable binding configuration can be induced 
by the resonant optical force.
\end{abstract} 

 ] %% activate for two-column option

\noindent Optical forces are useful in the manipulation of ultra-fine particles and 
mesoscopic systems, and the development is rather astounding in the last 
three decades. The most well known types of the optical 
forces are the radiation pressure and the optical gradient force. There is 
also an inter-particle optical force, induced by the multiple scattering of 
light.\cite{Lin:1,Burns:1989,Tatarkova:2002,and:2002,Antonoyiannakis:1997,
Singer:2003,Chaumet:2001} 
We present here an interesting type of resonant 
inter-particle force. We will see that the tuning of the incident light 
frequency to the Morphology Dependent Resonance (MDR) of a cluster of 
transparent microspheres would induce a strong resonant optical force 
(MDR-force) between the spheres. The MDR of a pair 
of spheres had been observed in fluorescent\cite{Mukaiyama:1999,Rakovich:2004} 
and lasing\cite{Hara:2003} experiments. Here we 
study theoretically the force induced by such resonances. We will see that 
the MDR-induced force, derived from the coherent coupling of the whispering 
gallery modes (WGM's), is a strong short ranged force that can be attractive 
or repulsive depending on whether the bonding mode (BM) or the anti-bonding mode (ABM) is 
excited. The strength of the optical forces can be enhanced by orders of 
magnitude when a MDR is excited. As microsphere cavities are emerging as an 
alternative to the photonic crystal in controlling light,\cite{Mukaiyama:1999,
Rakovich:2004,Hara:2003} the MDR-force may be 
deployed for the manipulation of a microsphere cluster.

In this paper, we calculate the electromagnetic (EM) forces acting on 
microspheres when WGM's or MDR's are excited. The optical force acting on a 
microsphere can be computed via a surface integral of the Maxwell stress 
tensor, $\mathord{\buildrel{\lower3pt\hbox{$\scriptscriptstyle\leftrightarrow$}}\over 
{T}} $, over the sphere's surface. The microspheres cannot respond to the 
high frequency component of the time varying optical force, so we calculate 
the time-averaged force $<\mathord{\buildrel{\lower3pt\hbox{$\scriptscriptstyle\rightharpoonup$}}\over 
{F}}>=\oint { <
\mathord{\buildrel{\lower3pt\hbox{$\scriptscriptstyle\leftrightarrow$}}\over 
{T}}> \cdot 
d\mathord{\buildrel{\lower3pt\hbox{$\scriptscriptstyle\rightharpoonup$}}\over 
{S}}}$. The EM field required in evaluating 
$\mathord{\buildrel{\lower3pt\hbox{$\scriptscriptstyle\leftrightarrow$}}\over 
{T}} $ is computed by the multiple scattering 
theory,\cite{Lin:1,Appl:1995} which expands the fields in vector spherical harmonics. This formalism is quite possibly the most accurate method that can be applied. It is in principle exact, and the numerical convergence is being controlled by the maximum angular momentum $(L_{max})$ used in the expansion. 
The calculation for the resonance of dielectric microspheres near contact 
requires a high $L_{max}$,\cite{Miyazaki:2000} which is chosen so that 
further increase in $L_{max}$ does not change the value of the calculated 
force. In most of the calculations, the size parameter (\textit{kR}) is between 28 and 
29, and $L_{max}$=63 was used. We adopt the Generalized Minimal Residual 
iterative solver (GMRES) for the linear system of 
equations.\cite{Fraysse:2003} In the following, the WGM's will be labeled 
as ``($l)$TE($n)$'' or ``($l)$TM($n)$'', where $l$ and $n$ are the mode and order number, and TE 
(TM) means transverse electric (magnetic) respectively. Unless otherwise 
noted, a linearly polarized incident plane wave with a modest intensity of 
10$^{4}$ W/cm$^{2}$ is assumed throughout this paper. The spheres have 
radius $R$=2.5 $\mu $m, with a dielectric constant \textit{$\varepsilon $}=2.5281+10$^{ - 4}i$. The 
loss level of \textit{Im}{\{}\textit{$\varepsilon $}{\}}=10$^{ - 4}$ or smaller can be easily achieved with 
insulators, glass or possibly good quality polystyrene spheres.

\begin{figure}[htbp]
\centerline{\includegraphics[width=3.26in,height=2.35in]{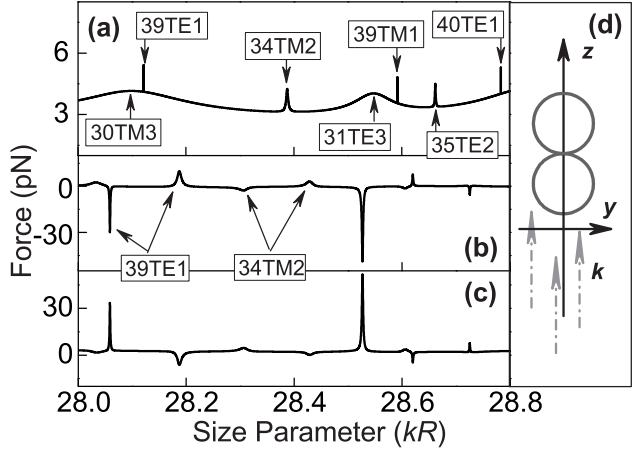}}
\caption{(a): The radiation pressure for a sphere with $\varepsilon 
$=2.5281. (b)-(c): Optical forces acting on two contiguous microspheres 
($\varepsilon $=2.5281+10$^{ - 4}i$), with configuration depicted in inset 
(d), with Panel (b) for the upper sphere and Panel (c) for the lower sphere. 
(d): A pair of contiguous spheres illuminated by a linearly polarized plane 
wave propagating along the bisphere ($z$) axis.}
\label{fig1}
\end{figure}

The well-known WGM's for a transparent microsphere have many interesting 
properties and applications, mostly because of its high quality factor and 
the enhanced EM fields near the surface. While the fields can be enhanced by 
orders of magnitude when a WGM is excited, the radiation pressure is only 
increased by about 30{\%} or less, as shown in Fig. \ref{fig1}(a). It is because the intensity distribution of a WGM is symmetrical, so 
that the gradient force acting on the sphere at any point is cancelled by 
its counterpart on the other side of the sphere. However, a much stronger 
enhancement in the optical force can be induced by the resonances involving 
two spheres. When two spheres are near each other, their EM modes are 
coherently coupled and split into BM's and ABM's through the quasi-normal mode 
splitting.\cite{Antonoyiannakis:1997,Miyazaki:2000} The 
BM's (ABM's) have resonant frequencies that are lower (higher) 
than that of the single sphere, and have an even (odd) parity in the EM field 
distribution.\cite{Miyazaki:2000} Unlike the single sphere resonance where 
the force is not enhanced that much, the MDR's correspond to strong 
attractions (BM's) or repulsions (ABM's) between the 
spheres. The overall intensity distribution of the two-sphere resonance is 
still symmetrical, but the field pattern on each sphere is not. The strong 
internal fields then induce strong optical forces on the spheres. We note 
that the BM and ABM forces are also observed between layers of 
2 dimensional photonic structure.\cite{Antonoyiannakis:1997}

In Fig. \ref{fig1}(b)-(c) we plot the optical forces acting 
on a pair of spheres with the geometry shown in Fig. \ref{fig1}(d). The wavelengths of the incident light fall inside the range of 542 nm 
to 561 nm, chosen to match with that of the previous works on 
MDR.\cite{Miyazaki:2000,Fuller:1991} The BM and 
ABM of 39TE1 and 34TM2 are marked on Fig. 
\ref{fig1}(b). When a resonance is excited, the force is 
tremendously enhanced compared to off-resonance. The BM's (ABM's) have the 
maximum (minimum) field intensity at the contact point of the spheres, giving rise to 
attractions (repulsions). The resonant linewidths of the MDR are 
also several orders of magnitude wider than that of a single 
sphere,\cite{Miyazaki:2000}$^{, }$\cite{Fuller:1991} and they are 
further broadened by absorption. We remark that the small peak at \textit{kR}=28.03 in 
Fig. \ref{fig1}(b)-(c) is the ABM of 34TE2, 
and also the interactions between 39TM1 and 35TE2 complicated the splitting, 
and their coupling give rise to the MDR-force peaks at \textit{kR}=28.527, 28.605 and 
28.620.

\begin{figure}[htbp]
\centerline{\includegraphics[width=2.5in]{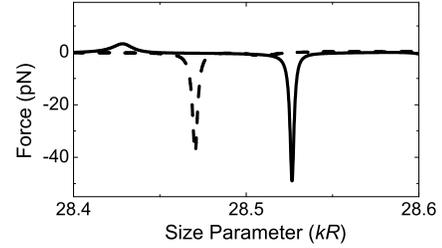}}
\caption{Optical forces acting on a pair of spheres with the configuration shown in 
Fig. \ref{fig1}(d). The horizontal axis is the size parameter 
of the bottom sphere. Solid lines: both spheres have radius of 2.5 $\mu $m. 
Dotted lines: The bottom sphere has radius of 2.5 $\mu $m and the top sphere 
has radius of 2.45 $\mu $m.}
\label{fig2}
\end{figure}

One of the major challenges in studying MDR of spheres experimentally is 
that the resonant frequency is very sensitive to the size of the sphere and 
thus requires extremely accurate particle sizing.\cite{Fuller:1991} This 
difficulty has been overcome by utilizing the narrow linewidth of the single 
sphere resonance to determine the particle size.\cite{Mukaiyama:1999,
Rakovich:2004,Hara:2003} Nevertheless, the MDR 
force is actually quite robust against size dispersion. The solid line in 
Fig. \ref{fig2} shows the MDR force at \textit{kR}=28.527 when the two spheres are of the same 
diameter, to be compared with the forces in which the two spheres differ by 
2{\%} in diameter (dotted line). We see that the MDR force remains 
significant even when the two spheres do not have the same radius.

\begin{figure}[htbp]
\centerline{\includegraphics[width=2.5in]{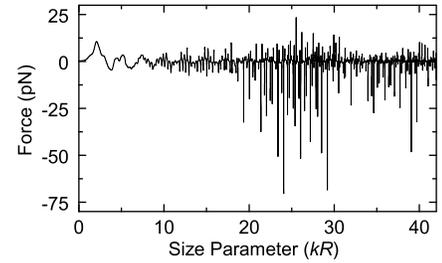}}
\caption{Optical forces as a function of the size parameter acting on two 
contiguous spheres as depicted in Fig. \ref{fig1}(d). 
$\varepsilon $=2.5281+10$^{ - 4}i$. Only the force acting on top sphere is 
plotted.}
\label{fig3}
\end{figure}

Figure \ref{fig3} shows the forces acting on a pair of 
spheres over a wide range of size parameters. From this figure, we see that 
the attractive resonant force is generally stronger than the repulsive 
resonant force. The resonant force is most significant for spheres with size 
parameters between 20 and 30. The force for those with size parameters 
greater than 30 is damped by absorption.

\begin{figure}[htbp]
\centerline{\includegraphics[width=3.18in,height=2.76in]{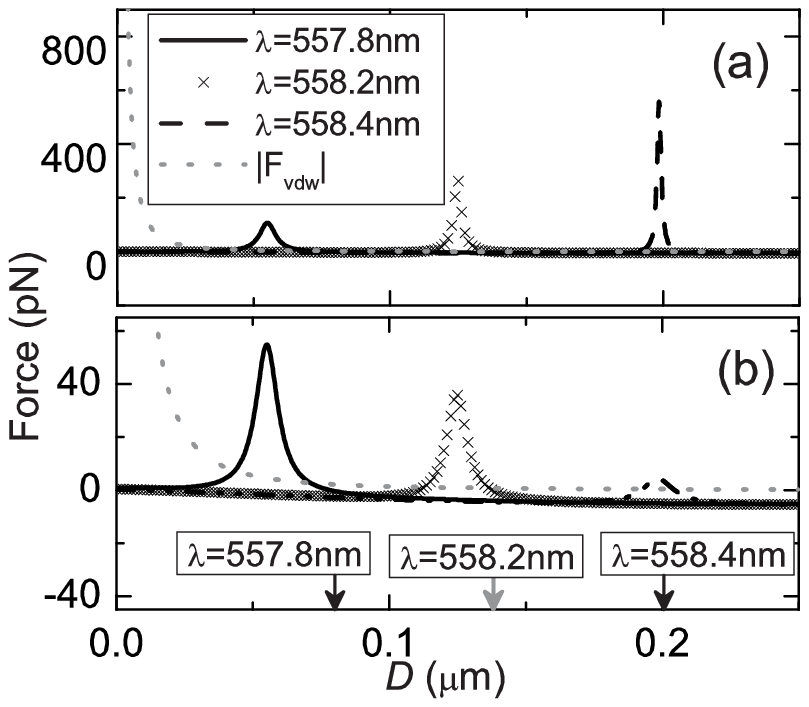}}
\caption{Optical force acting on a pair of spheres plotted as a function of $D$, the 
separation between the closest points on the spheres. The forces acting on 
the spheres are equal and opposite by symmetry, with positive force represents
 repulsion and vice versa. The positions of the spheres are 
$(0, 0, -D/2-R)$ and $(0, 0, D/2+ R)$. The incident wave has the form 
$\mathord{\buildrel{\lower3pt\hbox{$\scriptscriptstyle\rightharpoonup$}}\over 
{E}} _{in} = \hat {x}E_o \sin (kz)$. The 39TE1 resonance of a single sphere 
is at $\lambda $=558.6 nm. $\vert $F$_{vdw}\vert $ is an upper bound of 
the magnitude of the van der Waals force. (a): Ideal case with no 
absorption, i.e. $\varepsilon $=2.5281. (b) $\varepsilon $=2.5281+10$^{ - 
4}i$. The stable equilibrium separations (optical force equals zero and 
stable against perturbation) for different incident wavelength are marked by 
arrows.}
\label{fig4}
\end{figure}

The MDR frequencies actually depend on the distance between the spheres, and 
this property can be utilized to bind the spheres into a stable structure. 
As an illustrative example, we consider a pair of spheres (aligned along 
$z$-axis) illuminated by an incident field of the form 
$\mathord{\buildrel{\lower3pt\hbox{$\scriptscriptstyle\rightharpoonup$}}\over 
{E}} _{in} = \hat {x}E_o \sin (kz)$, which compose of a pair of 
counter-propagating waves going along the bisphere axis. At a particular 
frequency of the incident wave, slightly higher than the resonant frequency 
of a WGM, the ABM is excited at a particular distance between 
the spheres, leading to strong repulsion. However, at distances larger than 
that particular distance, both the radiation pressure and the Van der Waals 
forces will push the balls together. This competition between ABM 
resonant repulsion and other attractive forces lead to the stable position. 
Figure \ref{fig4} shows the force as a function of $D$, the 
separation between the closest points on the spheres. The dielectric 
constant is taken to be 2.5281+10$^{ - 4}i$ in 
Fig. \ref{fig4}(b), and the ideal case results with no 
absorption (\textit{$\varepsilon $}=2.5281) are shown in Fig. \ref{fig4}(a) for 
comparison. Stable equilibrium separations, where the optical force is zero, 
are marked by arrows in Fig. \ref{fig4}(b). The spheres will 
experience an attractive (repulsive) force if their separation is increased 
(decreased) from the equilibrium distance. Binding can also be achieved by 
using two lasers, one tuned to a BM and the other tuned to an 
ABM such that there is an equilibrium separation 
``sandwiched'' by the resonant force peaks. The interaction between the two 
laser beams can be neglected because of the lack of coherence.

We also compare the MDR-force with other relevant interactions. The energy 
associated with the repulsive barriers created by the ABM's are 
on the order of tens of $k_{b}T$ (the thermal energy at room temperature) at 
an incident intensity of 10$^{4}$ W/cm$^{2}$. For example, it takes about 
80 $k_{b}T$ to push the spheres across the middle peak of Fig. \ref{fig4}(b)
 (corresponding to $\lambda $=558.2 nm). Another relevant 
comparison is the strength of the van der Waals forces. An upper bound on 
the magnitude of the van der Waals force between two dielectric spheres, 
$\vert F_{vdw}\vert $, can be calculated by the non-retarded 
approximation: $\vert F_{vdw} (D)\vert \le AR / 12D^2$, where $A$=6.6$\times 
$10$^{ - 20}$ Joule is the Hamaker constant.\cite{See:1991} The magnitude 
of the van der Waals force is plotted on Fig. \ref{fig4}. One 
sees that the resonant force can dominate over the van der Waals force if 
the $D$ is more than a few tens of nano-meter. Finally, the weight of a glass 
sphere (mass density =2400 kg/m$^{3})$ is about 1.5 pN.

We note from Fig. \ref{fig4} that the resonant separation 
(where the force is maximum) increases as the incident frequency is tuned 
closer to the resonant frequency of the WGM. This can be understood from the 
fact that a larger separation corresponds to a smaller splitting of the WGM. 
In the ideal case with no absorption (see \ref{fig4}(a)), the strength of the MDR-force is an increasing function of resonant 
separation. This is because the quality factor, and thus the internal field 
of the MDR, attains the huge values of the WGM as the separation 
increases.\cite{Miyazaki:2000} We note that resonant force for the ideal 
case approaches a nano-Newton. However, in reality the resonances are 
inevitably subject to absorptive losses.

We emphasize that the properties of the resonant mode is determined by the 
morphology. As long as the incident frequency matches the resonant 
frequency, the resonance will be excited irrespective of the external light 
profile. However, it is the projection (coupling) of the incident light onto the 
resonating mode that determines the strength of the resonat force. A 
plane wave is in fact not the most efficient way to excite the MDR, as most 
of the light is coupled to the non-resonating, dissipative modes. Our calculations 
aim to illustrate the resonant behavior and the corresponding strong optical 
forces. In actual implementation, other form(s) of incident light wave (e.g. evanescent wave) can 
be used to realize a stronger force and thereby to utilize the full 
potential of the resonant effect. We also note that while absorption will 
degrade the strength of the resonance, microspheres containing gain 
materials can in principle enhance the resonant force, and the effect should 
be most interesting when the WGM starts lasing.\cite{Hara:2003} These 
would be interesting topics for further studies.

Support by Hong Kong RGC through CA02/03.SC05 and HKUST6138/00P is 
gratefully acknowledged. Zhifang Lin is also supported by CNKBRSF and NNSF 
of China. C.T. Chan's e-mail address is phchan@ust.hk.

\clearpage
\end{document}